\documentstyle[12pt,aasms4,epsfig]{article}

\lefthead{A.\ B.\ Kaye, et al.\ }
\righthead{$\gamma$ Doradus Stars}

\begin{document}

\title{$\gamma$ Doradus Stars:  Defining a New Class of Pulsating Variables}

\author{Anthony B.\ Kaye}
\affil{Applied Theoretical and Computational Physics Division, Los Alamos National Laboratory}
\authoraddr{X-TA, MS B-220, Los Alamos, New Mexico  87545  U.S.A.}

\author{Gerald Handler}
\affil{Institut f\"ur Astronomie, Universit\"at Wien}
\authoraddr{T\"urkenschanzstra\ss e 17, A-1180 Vienna, Austria}

\author{Kevin Krisciunas}
\affil{Department of Astronomy, University of Washington}
\authoraddr{Box 351580, Seattle, WA 98195-1580, USA}

\author{Ennio Poretti and Filippo M.\ Zerbi}
\affil{Osservatorio Astronomico di Brera/Merate}
\authoraddr{Via Bianchi 46, I-23807 Merate (Lc), Italy}

\begin{abstract}

In this paper we describe a new class of pulsating stars, the prototype 
of which is the bright, early, F-type dwarf, $\gamma$ Doradus.  These stars 
typically have between 1 and 5 periods ranging from 0.4 to 3 days with photometric 
amplitudes up to $0\fm 1$ in Johnson $V$.  The mechanism for these 
observed variations is high-order, low-degree, non-radial, gravity-mode pulsation.


\end{abstract}

\section{Introduction}

Cousins \& Warren (1963) discovered that the bright F0 {\sc v} star
$\gamma$ Doradus was variable over a range of several hundredths of a
magnitude with two principal periods ($0\fd 733$ and $0\fd 757$). 
$\gamma$ Doradus has an absolute magnitude similar to that of a 
$\delta$ Scuti star, but is somewhat cooler, and thus for many years 
it was deemed a ``variable without a cause".
Cousins (1992) stated: ``The suggested W-UMa type no longer seems a
possibility, but rotation with starspots and/or tidal distortion might
account for the variability. The light-curve and dual periodicity would
favor some form of pulsation, but the period is much longer than expected
for a $\delta$ Scuti star." Balona et al. (1994) tried to
model the star using two starspots and differential rotation. They found that
the large size of the required spots and the high stability of their
periods did not bode well for the starspot hypothesis. Furthermore, they found
evidence of a third period, later confirmed by Balona et al.\ (1996),
which further diminishes the likelihood of the starspot hypothesis.

9 Aurigae (= HD 32537), a star very similar to $\gamma$ Doradus, 
was first noted to be variable by Krisciunas \& Guinan (1990). 
Krisciunas et al.\ (1993) 
found evidence for two photometric periods between 1.2 and 3 days. 
Using infrared and IUE data, Krisciunas et al.\ (1993) found no evidence 
for a close companion or a lumpy ring of dust surrounding the star, but they could
not rule out the idea of starspots. 

Over the past decade, more than 40 variable stars with spectral types and
luminosity classes similar to $\gamma$ Doradus have been discovered that
exhibit variability on a time scale that is an order of magnitude slower 
than $\delta$ Scuti stars.  
Mantegazza et al.\ (1994), Krisciunas (1994), 
and Hall (1995) suggested that these objects may constitute a new class of variable stars.  
Breger \& Beichbuchner (1996) investigated whether any known $\delta$ Scuti 
stars also showed $\gamma$ Doradus-type behavior and found no clear cut 
examples of stars that show both ``fast'' and ``slow'' variability; Fig.\ 1 of 
their paper nicely illustrates the locations of the two kinds of variables in the 
color-magnitude diagram.  However, not all of their $\gamma$ Doradus stars 
are regarded as {\it bona fide} members of the group.

Krisciunas (1998) provides a good summary of our knowledge of $\gamma$ 
Doradus stars as a new class, but to date there is no publication in the 
refereed journal literature which 
summarizes and ``defines'' the characteristics of the class itself.  It was quite 
evident early on that significant advancement in the understanding of the 
physical nature of $\gamma$ Doradus stars could be made only on the basis of 
a large observational effort.  Hence, activities were concentrated
in international multi-longitude photometric and spectroscopic campaigns.

On the basis of extensive photometry, radial velocities, and line-profile 
variations, it has been proven that 9 Aurigae (Krisciunas et
al.\ 1995a, Zerbi et al. 1997a, Kaye 1998a), $\gamma$ Doradus
(Balona et al.\ 1996), HD 164615 (Zerbi et al.\ 1997b; Hatzes
1998), HR 8330 (Kaye et al.\ 1999), HD 62454 and HD 68192 (Kaye
1998a), and HR 8799 (Zerbi et al.\ 1999) are indeed pulsating variable stars. 
Given the nature of
the observed variability in these stars, 
the cause must be high-order ($n$), low-degree ($\ell$), 
non-radial $g$-modes.  We assert this on the basis of evidence 
{\em for} non-radial $g$-modes and the lack of convincing evidence for other 
explanations, including starspots.  Furthermore, 
we argue that since this small (but growing) group of objects 
all have similar physical characteristics 
and show broad-band light-- and line-profile variations resulting from the same 
physical mechanism, they form a new class of variable stars.  In this paper, we indicate 
the cohesiveness of this group and its differences from other variable star classes.  
Finally, we provide a set of criteria by which new candidates may be judged.


\section{General Characteristics of the Class}

Our list of {\it bona fide} $\gamma$ Doradus stars is complete to April 1999
and all objects of this class have extensive enough photometric and/or 
spectroscopic data sets to rule out other variability mechanisms.  
A complete, commented, up-to-date list of all proposed candidates for this 
group, as well as their observational history, is kept by Handler and 
Krisciunas at the World Wide Web site: 
{\it http://www.astro.univie.ac.at/$\sim$gerald/gdor.html}.

Table 1 lists the observed quantities of each of the 13 objects used to define
this new class of variable stars. Column 1 gives the most common name of each object.
Column 2 provides the best available value of $(b-y)$; columns
3 and 4 list the average apparent visual magnitude of each object ($<V>$) and the
best determined spectral type.  Column 5 lists the best available value of the 
projected equatorial velocity, $v \sin i$, in km s$^{-1}$. Column 6 reports 
the {\sc hipparcos} trigonometric parallax in milli-arcseconds (ESA 1997).

\begin{table}
\caption{Observational Parameters of the Confirmed $\gamma$ Doradus Variables}
\begin{tabular}{ccccccc}
\hline
Star & $(b-y)$ & $<V>$ & Spectral & $v \sin i$ & $\pi$ & Principal \\
 & (mag) &  (mag) & Type & (km s$^{-1}$) & (mas) & Reference \\
\hline
HD 224945 & 0.192 & 6.93 & F0$+$ {\sc v}$^{c}$ &  55 & 16.92 & 1\\
$\gamma$ Dor & 0.201 & 4.25 & F0 {\sc v} &  62 & 49.26 & 2\\
9 Aur & 0.217 & 5.00 & F0 {\sc v} &  18 & 38.14 & 3\\
BS 2740 & 0.219 & 4.49 & F0 {\sc v} &  40 & 47.22 & 4\\
HD 62454 & 0.214 & 7.15 & F1 {\sc v}$^{a}$ &  53 & 11.18 & 5\\
HD 68192 & 0.227 & 7.16 & F2 {\sc v}$^{c}$ & 85 & 10.67 & 5\\
HD 108100 & 0.234 & 7.14 & F2 {\sc v} & 68 & 12.10 & 6\\
BS 6277 & 0.167 & 6.20 & F0 {\sc v} & 185 & 13.70 & 7 \\
HD 164615 & 0.226 & 7.06 & F2 {\sc iv}$^{c}$  & 66 & 14.36 & 8\\
BS 6767 & 0.183 & 6.40 & F0 {\sc v}n$^{c}$ & 135 & 17.44 & 5\\
BS 8330 & 0.225 & 6.20 & F2 {\sc iv}$^{c}$ &  38 & 19.90 & 8\\
BS 8799 & 0.181 & 5.99 & kA5 hF0 mA5 {\sc v}; $\lambda$ Boo$^{b}$ &  45 & 25.04 & 9\\
HD 224638 & 0.198 & 7.49 & F1 {\sc v}s$^{c}$ &  24 & 12.56 & 10\\
\hline
\end{tabular}

\medskip

{\bf References:} (1) Poretti et al.\ 1996; (2) Balona, Krisciunas, \& Cousins 1994;
(3) Zerbi et al.\ 1997a; (4) Poretti et al.\ 1997;
(5) Kaye 1998a; (6) Breger et al.\ 1997;
(7) Zerbi et al.\ 1997b; (8) Kaye et al.\ 1999;
(9) Zerbi et al.\ 1999; (10) Mantegazza, Poretti, \& Zerbi 1994.

\medskip

$^{a}$HD 62454 is the primary star of a double-lined spectroscopic binary.  See 
Kaye 1998b. 

\smallskip

$^{b}$See Gray \& Kaye 1999a.

\smallskip

$^{c}$See Gray \& Kaye 1999b.

\end{table}

Table 2 presents derived properties of the thirteen objects.  
Estimates for the total metallicity ($[Me/H]$) 
are derived from the relations of Nissen (1988) and Smalley (1993), 
which are precise to within 0.1 dex in $[Me/H]$ and 
are listed in Column 2. 
The absolute visual magnitudes (Column 3) are calculated from the {\sc hipparcos} 
parallaxes. Luminosities, using
bolometric corrections listed in Lang (1992) and $M_{{\rm bol}, \sun}=4.75$
(Allen 1973), are presented in Column 4. 
The effective temperatures are determined from the new calibration of
Str\"{o}mgren photometry by Villa (1998), for which we estimate errors 
of $\pm$~100~K (Column 5); stellar radii precise to $\pm~0.05~R_{\sun}$ 
are then calculated (Column 6). 
Finally, masses which are precise to $\pm~0.03~M_{\sun}$ (internal model error), 
are inferred by comparison with solar-metallicity 
evolutionary tracks by Pamyatnykh et al.\ (1998) (Column 7).  The final entry
in Table 2 represents the unweighted average of each of the columns;
presumably, these are the physical parameters of a ``typical'' $\gamma$ Doradus variable.

\begin{table}
\caption{Calculated and Inferred Basic Properties of the Confirmed $\gamma$ Doradus 
Variables}
\begin{tabular}{ccccccc}
\hline
Star & [$Me/H$] & $M_V$ & $L/L_{\sun}$ & $T_{\rm eff}$ & $R/R_{\sun}$ & $M/M_{\sun}$ \\
 & & (mag) & & (K) & &  \\
\hline
HD 224945 & $-0.30$ & 3.07 & 5.1 & 7250 & 1.43 & 1.51 \\
$\gamma$ Dor & $-0.02$ & 2.72 & 7.0 & 7200 & 1.70 & 1.57 \\
9 Aur & $-0.19$ & 2.89 & 6.0 & 7100 & 1.62 & 1.52 \\
BS 2740 & $-0.15$ & 2.86 & 6.2 & 7100 & 1.64 & 1.53 \\
HD 62454 &  0.16 & 2.39 & 9.5 & 7125 & 2.02 & 1.66 \\
HD 68192 & 0.05 & 2.30 & 10.5 & 7000 & 2.20 & 1.71 \\
HD 108100 &  $-0.03$ & 2.53 & 8.5 & 6950 & 2.01 & 1.62 \\
BS 6277 & 0.09 & 1.93 & 14.7 & 7350 & 2.36 & 1.84 \\
HD 164615 &  0.20 & 2.82 & 6.5 & 7000 & 1.73 & 1.53 \\
BS 6767 &  $-0.10$ & 2.59 & 7.9 & 7300 & 1.76 & 1.61 \\
BS 8330 & $-0.01$ & 2.67 & 7.4 & 7000 & 1.85 & 1.57 \\
BS 8799 &  $-0.36$ & 2.96 & 5.7 & 7375 & 1.46 & 1.54 \\
HD 224638 &  $-0.15$ & 2.98 & 5.5 & 7200 & 1.51 & 1.52 \\
\hline
``Average'' & $-0.06$ & 2.69 & 7.6 & 7160 & 1.77 & 1.59 \\
\hline
\end{tabular}
\end{table}

We present a color-magnitude diagram of all 13 stars, using the {\sc hipparcos}
parallaxes to calculate accurate values of $M_{V}$ in Figure 1. 
The observed zero-age main sequence 
(Crawford 1975) and the observed edges of the $\delta$ Scuti instability 
strip (Breger 1979) are shown as a solid line and dashed lines, respectively.

\begin{figure}
\centerline{\psfig{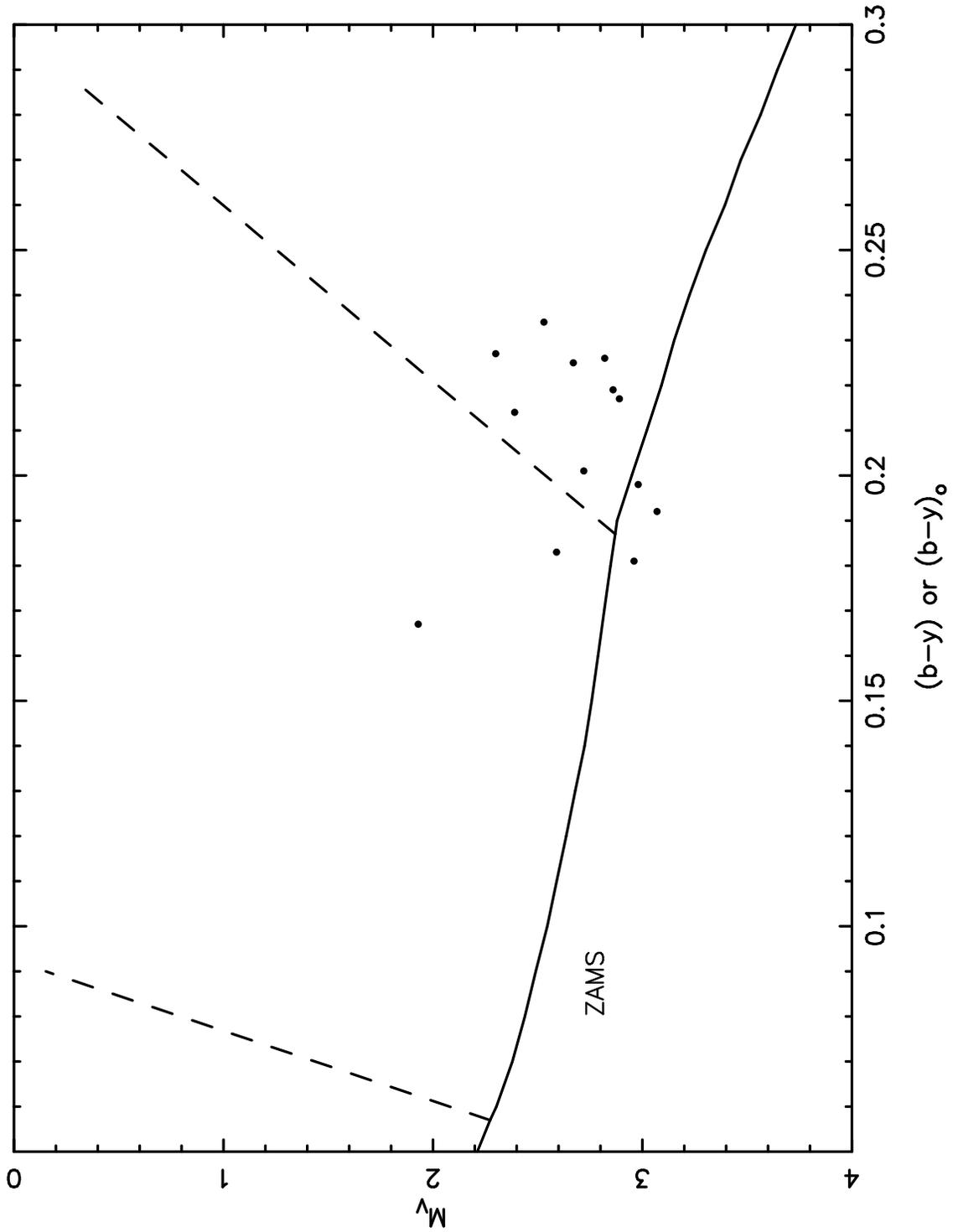}}
\caption{A color-magnitude diagram showing the positions of the confirmed 
$\gamma$ Doradus stars. Solid points indicate the position of each object, the
solid line represents the observed ZAMS, and the boundaries
of the observed $\delta$ Scuti instability strip are indicated by dashed lines.}
\end{figure}

The truly intriguing characteristic of $\gamma$ Doradus stars is that
they are variable; considering the part of the Hertzsprung-Russell 
diagram in which
they lie, previous pulsational models say they should not be.  The outer 
convection zones of these stars are too shallow to generate and sustain 
a large magnetic dynamo, thus making starspots improbable.  
Most of the $\gamma$ Doradus stars are multi-periodic; the average 
period is close to 0.8 days. The observed variations 
are not necessarily stable, and may be highly dynamic (Kaye \& Zerbi 1997).
Typical amplitudes cluster around 4 percent (= $0\fm 04$) in
Johnson $V$, and may vary during the course of an observing season
by as much as a factor of four.  For the best-studied stars (e.g., 
$\gamma$ Doradus itself, 9 Aurigae, and HR 8330), line-profile variations 
with periods equal to the photometric periods have been confirmed
(Balona et al.\ 1996; Kaye 1998a; Kaye et al.\ 1999).  No
high-frequency signals have been detected in either the 
photometry or the spectroscopy, indicating a lack of the $p$-mode
pulsation common in $\delta$ Scuti stars.

Despite their commonality, a small subset of $\gamma$ Doradus stars 
show remarkably peculiar pulsation characteristics.  In 
several objects [e.g., HD 224945 (Poretti et al.\ 1998), HD 224638 (Poretti
et al. 1994), and 9 Aurigae (see Krisciunas et al.\ 1995, Zerbi 
et al.\ 1997a, Kaye 1998)], amplitude variability of order 50\% over a few 
years is observed.  Other objects [e.g., $\gamma$ Doradus (Cousins 1992), HD 164615 
(Zerbi et al.\ 1997b), and HR 8799 (Zerbi et al.\ 1999)] show amplitude modulation 
selectively located at the moment of {\it maximum} brightness, a characteristic of variability 
that is new to the field of stellar pulsation.  Still other objects 
(e.g., HD 68192) show remarkably constant periods and amplitudes over long time scales.  
Clearly, these peculiarities within the $\gamma$ Doradus class need many more long-term 
observations to be explained.


\section{Defining a New Class}

We argue that the qualities and characteristics of the thirteen (13) above named 
and described stars form a homogeneous set based on their physical 
characteristics and their mechanism for variability, and thus form the basis for a new 
class of variable stars.

In following with the informal discussions at the
``Astrophysical Applications of Stellar Pulsation" conference
(Stobie \& Whitelock 1995) held in 1995 at Cape Town, South Africa and in 
recent papers in the literature (see e.g., Krisciunas et al.\ 1993,  
Balona et al.\ 1996, Zerbi et al.\ 1997a, Poretti et al.\ 1997,
Kaye 1998a, Kaye et al.\ 1999), we propose that this type of
variable star henceforth be known and recognized by the name 
{\it $\gamma$ Doradus variable stars}.  The extent of the $\gamma$ Doradus 
phenomenon, as it is currently known, consists of variable stars with an implied 
range in spectral type 
A7--F5 and in luminosity class {\sc iv}, {\sc iv-v}, or {\sc v}; their 
variations are consistent with the model of high-order ($n$), low-degree 
($\ell$), non-radial, gravity-mode oscillations.  Although it is conceivable that 
variations such as those of the stars in this class may occur outside of this 
region, it is likely that other mechanisms of variability would then dominate, and 
thus this combination of spectral type, luminosity class, and (most importantly) 
variability mechanism, forms a suitable definition.  

From an observational point of view, the $g$-mode oscillations seen in
$\gamma$ Doradus variables are characterized by periods between 
0.4 and 3 days and peak-to-peak amplitudes $\la 0\fm 1$ in Johnson $V$.
The presence of multiple periods and/or amplitude modulation is common among 
these stars, but is not included in the formal definition presented here.  
Spectroscopic variations are also observed, and these manifest themselves
both as low-amplitude radial-velocity variations (that cannot 
be attributed to duplicity effects) and as photospheric 
line-profile variations. 

In addition to these features, we stress that any object put forth for
consideration as a confirmed $\gamma$ Doradus variable star
must not vary {\it exclusively} by other mechanisms, including: $p$-mode 
pulsations (e.g., $\delta$ Scuti stars), 
rotational modulation of dark, cool,
magnetically-generated starspots; rotational modulation of bright, hot, 
abundance-anomaly regions; duplicity-induced variations; or other rotational effects.  
Obviously, dual-nature objects (e.g., pulsating stars showing both $\gamma$ Doradus-- 
and $\delta$ Scuti-type behavior) must not be rejected.  
Prime candidates for $\gamma$ Doradus stars should therefore {\it not} be 
primarily variable due to the rotational modulation occuring in Am stars, 
Ap stars, Fm stars, RS CVn stars, or BY Dra stars. However, candidates 
{\it may} be members of a spectroscopically-defined class (e.g., $\lambda$ Bo\"{o}tis stars; 
see, e.g., Gray \& Kaye 1999a).

\section{Concluding Perspective}

$\gamma$ Doradus stars constitute a new class of variable stars because they
all have about the same mass, temperature, luminosity, and the same mechanism of variability.  
They are clearly not a sub-class of any of the other A/F-type variable or peculiar 
stars in this part of the HR diagram, and may offer additional insight into stellar physics 
when they are better understood (e.g., they may represent the cool portion of an 
``iron opacity instability strip" currently formed by the $\beta$ Cephei stars, the 
SPB stars, and the subdwarf B stars; they may also 
offer insight into the presence of $g$-modes in solar-like stars).  Modeling by 
Kaye et al. (1999) is beginning to shed light on the 
theoretically required interior structure and on the specific physics driving 
the observed variability, but much theoretical work lies ahead.

To understand the behavior of $\gamma$ Doradus stars and to investigate how they
differ from the $\delta$ Scuti variables and spotted stars, we need to investigate a number of
star clusters of differing ages, perhaps up to as old as 1 Gyr.  
The fact that the Hyades has no $\gamma$ Doradus variables (Krisciunas et al.\ 1995b) 
may be a quirk of the Hyades, rather than proof that stars $\approx$ 600 Myr old 
are too old to exhibit $\gamma$ Doradus-type behavior.  Clearly, the ``outliers'' 
of the $\gamma$ Doradus candidates that would extend the limits of the region of 
the HRD in which these new variables are found should be checked carefully 
for both photometric and spectroscopic evidence indicative of pulsations 
versus starspots, duplicity effects, and other causes of variability 
not consistent with the definition presented above (see, e.g., Aerts et al.\ 1998).  
Finally, additional observations of 
individual $\gamma$ Doradus stars are clearly warranted in order to 
understand better the nature of these objects.  After all, thirteen objects does not 
an instability strip make.  In the meantime, we must keep an open and critical mind 
about these variables.

\section{Acknowledgments}

This work was performed under the auspices of the U.\ S.\ Department of 
Energy by the Los Alamos National Laboratory under contract No.\ W-7405-Eng-36.  
We gratefully acknowledge the unpublished spectral types of some of the stars in 
this paper from R.\ O.\ Gray, and we thank Holger Pikall for computing evolutionary 
models upon request.  ABK also gratefully acknowledges Drs.\ Guzik and Bradley 
for reading various drafts of this paper.  GH was partially supported by the Austrian 
Fonds zur F\"{o}rderung der wissenschaftlichen Forschung under grant No.\ S7304-AST.


\begin{thebibliography}{}


\bibitem[]{} Aerts C., Eyer L., Kestens E.  1998, A\&A, 337, 790
\bibitem[]{} Allen C.W. 1973, in Astrophysical Quantities (3rd ed., 
London: Athlone Press), p. 161, ff.
\bibitem[]{} Balona L.A., Krisciunas K., Cousins A.W.J. 1994,
MNRAS, 270, 905
\bibitem[]{} Balona L.A., $\rm B\ddot{o}hm$ T., Foing B. H., et al. 
1996, MNRAS, 281, 1315
\bibitem[]{} Breger M. 1979, PASP, 91, 5
\bibitem[]{} Breger, M.\ \& Beichbuchner, F., 1996, A\&A, 313, 851
\bibitem[]{} Breger M., Handler G., Garrido R., et al., 1997, A\&A, 
324, 566
\bibitem[]{} Cousins A.W.J. 1992, Observatory, 112, 53
\bibitem[]{} Cousins A.W.J., Warren P. R. 1963, Mon. Notes Astron. 
Soc. S. Afr., 22, 65
\bibitem[]{} Crawford D.L.  1975, AJ, 80, 955
\bibitem[]{} ESA 1997, The {\sc hipparcos} Catalogue, Publ. SP-1200 (ESA,
Paris)
\bibitem[]{} Gray R.O., Kaye A.B., 1999a, AJ, submitted
\bibitem[]{} Gray R.O., Kaye A.B., 1999b, PASP, in preparation
\bibitem[]{} Hall D.S. 1995, in Robotic Telescopes: Current
Capabilities,  Present Developments, and Future Prospects for Automated
Astronomy, ed. G. W. Henry \& J. A. Eaton, ASP Conf. Ser. Vol. 79 (San
Francisco: ASP), p. 65
\bibitem[]{} Hatzes A.P. 1998, MNRAS, 299, 403
\bibitem[]{} Kaye A.B., Guzik J.A., \& Bradley P.A.  1999, in press
\bibitem[]{} Kaye A.B. 1998a, Ph.D.\ Thesis, Georgia State Univ.
\bibitem[]{} Kaye A.B. 1998b, IBVS No.\ 4596
\bibitem[]{} Kaye A.B., Henry G.W., Fekel F.C., Hall D.S. 1999, 
MNRAS, in press
\bibitem[]{} Kaye A.B., Zerbi F.M.  1997, DSSN, 11, 32
\bibitem[]{} Krisciunas K., Guinan E.F. 1990, IBVS 3511
\bibitem[]{} Krisciunas K., et al. 1993, MNRAS, 263, 781
\bibitem[]{} Krisciunas K. 1994, Comments Astrophys., 17, 213
\bibitem[]{} Krisciunas K., Griffin R. F., Guinan E. F., Luedeke K. D., 
McCook G. P. 1995a, MNRAS, 273, 662
\bibitem[]{} Krisciunas K., Crowe R. A., Luedeke K. D., Roberts M.,
1995b, MNRAS 277, 1404
\bibitem[]{} Krisciunas, K., 1998, in New Eyes to See Inside the Stars, ed.\ F.L.\ Deubner, 
IAU Symp.\ 185 (Dordrecht: Kluwer), p.\ 339
\bibitem[]{} Lang K. 1992, Astrophysical data: planets and stars
(New York: Springer), p. 138
\bibitem[]{} Mantegazza L., Poretti E., Zerbi F.M. 1994, MNRAS, 270 439
\bibitem[]{} Nissen P.E. 1988, A\&A, 199, 146
\bibitem[]{} Pamyatnykh A.A., Dziembowski W.A., Handler G., Pikall, H.
1998, A\&A, 333, 141
\bibitem[]{} Poretti E., Akan C., Bossi M., et al.,
Proc. IAU Coll.\ 181, ''Sounding Solar and Stellar Interiors'', Poster Volume,
ed. J.\ Provost \& F.X.\ Schmider, p.\ 279
\bibitem[]{} Poretti E., Koen C., Martinez P., et al., 1997, MNRAS, 292, 621
\bibitem[]{} Rodr\'{\i}guez E., Zerbi F. M., 1995, IBVS, No. 4170
\bibitem[]{} Smalley B. 1993, A\&A, 274, 391
\bibitem[]{} Stobie R.S., \& Whitelock P.A., eds. 1995, 
Astrophysical Applications of Stellar Pulsation, 
ASP Conf. Ser. Vol. 83 (Provo: ASP)
\bibitem[]{} Villa P. 1998, M.Sc.\ Thesis, University of Vienna
\bibitem[]{} Zerbi F.M., Garrido R., Rodr\'{\i}guez E., et al. 1997a, 
MNRAS, 290, 401
\bibitem[]{} Zerbi F.M., Rodr\'{\i}guez E., Garrido R., et al. 1997b,
MNRAS, 292, 43
\bibitem[]{} Zerbi F.M., Rodr\'{\i}guez E., Garrido R., et al.  1999, MNRAS, 303, 275

\end{thebibliography}
\end{document}